\documentclass[prd,twocolumn,showpacs,preprintnumbers,superscriptaddress,floatfix,nofootinbib]{revtex4-1}
\usepackage{subfigure}
\usepackage{amsfonts}
\usepackage{enumitem}
\usepackage{graphicx,,booktabs}
\usepackage{amssymb}
\usepackage{amsmath,txfonts,ulem}
\usepackage{graphicx}
\usepackage{dcolumn}
\usepackage{multirow}
\usepackage{color}
\usepackage{bm}
\usepackage{ulem}
\usepackage{makecell}
\usepackage[colorlinks,
citecolor=blue,
anchorcolor=green,
menucolor=orange,
linkcolor=red,
filecolor=red,
runcolor=pink,
urlcolor=blue,
frenchlinks=red]{hyperref}

\allowdisplaybreaks[4]

\begin{document}

\title{Analysis of strong coupling constant  with machine learning and its application}
	\author{Xiao-Yun Wang}
	\email{xywang@lut.edu.cn}
	\affiliation{Department of physics, Lanzhou University of Technology,
		Lanzhou 730050, China}
	\affiliation{Lanzhou Center for Theoretical Physics, Key Laboratory of Theoretical Physics of Gansu Province, Lanzhou University, Lanzhou, Gansu 730000, China}
	
	\author{Chen Dong}
	\affiliation{Department of physics, Lanzhou University of Technology,
		Lanzhou 730050, China}
	
	\author{Xiang Liu}
\email{xiangliu@lzu.edu.cn (Corresponding author)}
\affiliation{School of Physical Science and Technology, Lanzhou University, Lanzhou 730000, China}
\affiliation{Research Center for Hadron and CSR Physics, Lanzhou University and Institute of Modern Physics of CAS, Lanzhou 730000, China}
\affiliation{Lanzhou Center for Theoretical Physics, Key Laboratory of Theoretical Physics of Gansu Province, Lanzhou University, Lanzhou, Gansu 730000, China}
\affiliation{MoE Frontiers Science Center for Rare Isotopes, Lanzhou University, Lanzhou 730000, China}
\affiliation{Key Laboratory of Quantum Theory and Applications of MoE, Lanzhou University,
Lanzhou 730000, China}

\begin{abstract}
In this work, we investigate the nature of the strong coupling constant and related physics. Through the analysis of accumulated experimental data from around the world, we employ the ability of machine learning to unravel its physical laws. The result of our efforts is a formula that captures the expansive panorama of the distribution of the strong coupling constant across the entire energy range. Importantly, this newly derived expression is very similar to the formula derived from the Dyson-Schwinger
equations based on the framework of Yang-Mills theory. By introducing the Euler number, $e$, into the functional formula of the strong coupling constant at high energies, we have successfully solved the puzzle of the infrared divergence, which allows for a seamless transition of the strong coupling constant from the perturbative to the non-perturbative energy regime. Moreover, the obtained ghost and gluon dressing function distribution results confirm that the obtained strong coupling constant formula can well describe the physical properties of the non-perturbed regime.
In addition, we investigate the QCD strong coupling constant result of the Bjorken sum rule $\Gamma_1^{p-n}$ and the quark-quark static energy $E_0(r)$, and find that the global energy scale can effectively interpret the experimental data.
The results presented in this work shed light on the puzzling properties of quantum chromodynamics and the intricate interplay of strong coupling constants at both low and high energy scales.

\end{abstract}

\maketitle


\section{Introduction}
The precise value of the strong coupling constant $\alpha_s$ is a crucial parameter in the study of quantum chromodynamics (QCD)  \cite{Deur:2023dzc}. It represents the strength of the interaction between quarks and gluons inside hadrons \cite{Wang:2022uch,Wang:2022zwz} and is a fundamental issue in the theory of strong interaction, which involves quarks and gluons. At high energies $Q$, known as the short-distance scale, the perturbative method is generally employed to solve the problem. Experimental measurements of $\alpha_s$ can be made in deep inelastic scattering experiments \cite{Blumlein:1994kw} due to the property of asymptotic freedom. However, at low energies $Q$, also known as the long-distance scale, the interaction between quarks and gluons becomes so strong that experimental measurements are challenging, and the results' uncertainty is significant. To address this problem, non-perturbative theory should be considered. To date, the exact value of $\alpha_s$, which varies with $Q$, remains a subject of ongoing research.

During the 1970s, a significant breakthrough was achieved by David J. Gross, Frank Wilczek \cite{Gross:1973id}, and H. David Politzer \cite{DavidPolitzer} in establishing a fundamental relationship between the strong coupling constant $\alpha_s(Q)$ and the energy scale $Q$. Their work was based on the concept of asymptotic freedom, which is widely regarded as the cornerstone of QCD theory. In recognition of their contributions, they were honored with the Nobel Prize in Physics in 2004 \cite{GrossWIL}. Their pivotal achievement involved deriving the functional behavior of the first-order renormalization equation (FR) $\alpha_s(Q)$ at the short-distance scale, elucidating its relationship with the energy scale $Q$
\cite{Weinberg},
\begin{equation}
	\alpha_s^{FR}(Q) = \frac{g^2(Q)}{4\pi} = \frac{12 \pi}{(33 - 2 n_f)\ln(Q^2/\Lambda^2)}, \label{eq:1}
\end{equation}
where $n_f$ denotes the flavor of the quarks and $\Lambda$ is the QCD scale parameter. This progress provides valuable insights into the intricate dynamics of strong interactions and remains a key foundation in our understanding of QCD.
Since the 1970s, physicists have made remarkable progress in the study of the strong coupling constant at short-distance scales through the development of various theoretical models and experiments.

The study of the interaction between quarks and gluons at low energies $Q$ (long distance) is particularly challenging due to the phenomenon of quark confinement. This confinement effect renders the perturbative approach inadequate, leading to exceedingly intricate calculations \cite{Deur:2016tte}. Consequently, a multitude of non-perturbative theoretical models have been developed to tackle this complexity, including these methods such as the effective charge approach, QCD spectral sum rules, holographic light-front QCD (HLF-QCD), Dyson-Schwinger equations (DSE), and analytic coupling \cite{Grunberg:1980ja, Grunberg:1982fw,Deur:2017cvd,Brodsky:1997de,Brodsky:2010ur,Gribov:1977wm,Zwanziger:1981kg,Zwanziger:1982na,Shirkov:1996cd,Stefanis:2009kv,Alekseev:2002zn, Narison:2018dcr,Braaten:1991qm}.
With continued exploration, various experiments \cite{Deur:2008rf,Deur:2022msf,Deur:2016tte,Deur:2005cf,Deur:2008rf,Deur:2021klh,Kim:1998kia,HERMES:1997hjr,HERMES:1998pau,HERMES:1998cbu,HERMES:2002mes,HERMES:2006jyl,Yu:2021yvw,Narison:2018dcr,Braaten:1991qm} have been developed to investigate the behavior of the strong coupling constant at low-energy scale. Nevertheless, due to the inherent complexity of this regime, differences persist among different theoretical approaches, contributing to discrepancies in the predictions.

The Bjorken sum rule (BSR) \cite{Bjorken:1966jh,Bjorken:1969mm} is a spin sum rule based on quantum chromodynamics that relates the nucleon axial charge $g_A$ to the integral of the spin structure functions of protons and neutrons. This sum rule not only reveals critical information about the spin structure of nucleons, but also is an effective tool to test the theory of strong coupling constant $\alpha_s(Q)$ \cite{Deur:2009zy,Deur:2009tj}. By utilizing the relationship between the Bjorken sum rule and the Gerasimov-Drell-Hearn (GDH) sum rule, a reliable and robust coupling constant known as $\alpha_{eff}$ can be established \cite{Ayala:2018ulm}. However, it should be noted that this method's effectiveness is limited in the small $Q^2$ range and is most suitable for the large $Q^2$ region. BSR has accumulated a large amount of experimental data in the experiment \cite{Deur:2023dzc,E143:1998hbs,Deur:2004ti,Deur:2008ej} and has established numerous theoretical models \cite{Stein:1995si,Balitsky:1989jb,Sidorov:2006vu,Deur:2021klh}. However, reasonable experimental verification is obtained only in large $Q^2$, and there are still significant theoretical uncertainties in small $Q^2$. Therefore, the valid interpretation of BSR should be based on the precise determination of the strong coupling constant $\alpha_s(Q)$.

The QCD static energy $E_0(r)$ \cite{Ananthanarayan:2020umo} is a crucial physical value frequently utilized to explain strong interactions \cite{Tormo:2013tha}. It indicates the energy of the connection between two stationary quarks. The static energy of QCD is related to the color charge and distance of quarks, and the strong coupling constant in the high-energy region is small. The $E_0(r)$ can be calculated by perturbation expansion \cite{Bazavov:2011nk,Necco:2001xg}. QCD static energy is a critical theory to study the structure and properties of hadrons, which can be used to determine the mass, color charge, and spin of quarks and gluons. In addition, an essential aspect of QCD static energy is to determine the QCD strong coupling constant \cite{Bazavov:2012ka,Michael:1992nj,Tormo:2013tha}. Similarly, the distribution of QCD static energy can also be verified based on the strong coupling constant.

To date, an accurate theoretical model that fully encompasses the behavior of the strong coupling constant at a global energy scale remains elusive. This situation parallels the historical challenges encountered in understanding electromagnetism \cite{ampere,Oersted,Maxwell,Maxwell2} and blackbody radiation \cite{Plank,Boltzmann}.


In recent years, with the continuous advancement of various algorithms, particularly artificial neural networks (ANN) and regression algorithms, which are fundamental in the field of artificial intelligence, there has been rapid progress in the development of integrated algorithms. Notably, AI Poincar$\grave{e}$ \cite{Liu:2020omw,Liu:2021azq1}, AI Feynman \cite{Udrescu:2019mnk,Udrescu:2019mnk2}, and $\Phi$-SO \cite{Tenachi2023} are noteworthy examples. These algorithms, based on popular frameworks such as TensorFlow or PyTorch \cite{Steven}, offer significant advantages in applying machine learning techniques to tackle physical problems.
For instance, a team at MIT harnessed the power of these algorithmic frameworks, specifically AI Poincar$\grave{e}$ and AI Feynman, to automate the discovery and recovery of conserved quantities \cite{Liu:2020omw,Liu:2021azq1}, as well as hidden symmetries \cite{Liu:2021azq}. This notable achievement demonstrates the potential of machine learning methods in addressing intricate physics-related challenges.
Furthermore, researchers at the University of Strasbourg have combined a recurrent neural network (RNN) with a symbolic regression algorithm to create a comprehensive algorithmic framework called $\Phi$-SO \cite{Tenachi2023}. This framework primarily utilizes recurrent neural networks and reinforcement learning strategies to identify and analyze optimal physical expressions of symbolic data and unit rules.
Undoubtedly, these integrated algorithm frameworks offer tremendous potential for the study of physical problems using machine learning approach.

In this work, we present an  application of the $\Phi$-SO algorithmic framework \cite{Tenachi2023} to perform machine learning research on the intricate functional relationship that governs the strong coupling constant over a global energy scale. This study represents the first implementation of the algorithmic framework in this context, and we believe it serves as a significant advance and exemplary demonstration of the potential of machine learning to address concrete physics problems, particularly in the field of particle physics. At the same time, the experimental data of the BSR and the QCD quark-quark static energy $E_0(r)$ are used to further verify the result of strong coupling constant generated by machine learning, proving the effectiveness of the  machine learning method in understanding the QCD strong interaction.
\section{Strong Coupling Constant}
Currently, QCD strong coupling constants are investigated employing perturbation and non-perturbation theories for high and low energy scales, respectively. However, coordinating these two theories on the global scale poses challenges and limitations. To tackle this challenge, we attempt to incorporate machine learning (ML) techniques to analyze the law of QCD, strong coupling constant $\alpha_s(Q)$. Machine learning methods are adaptive and flexible and can learn valuable information from data without relying on a specific theoretical model. We analyze experimental data of the QCD strong coupling constant at various energy scales with ML. This helps us understand the trend and characteristics of its variation, leading to new ideas and techniques for comprehending the physical nature of the $\alpha_s(Q)$.

\begin{figure}[tbp]
	\centering
	\includegraphics[scale=0.35]{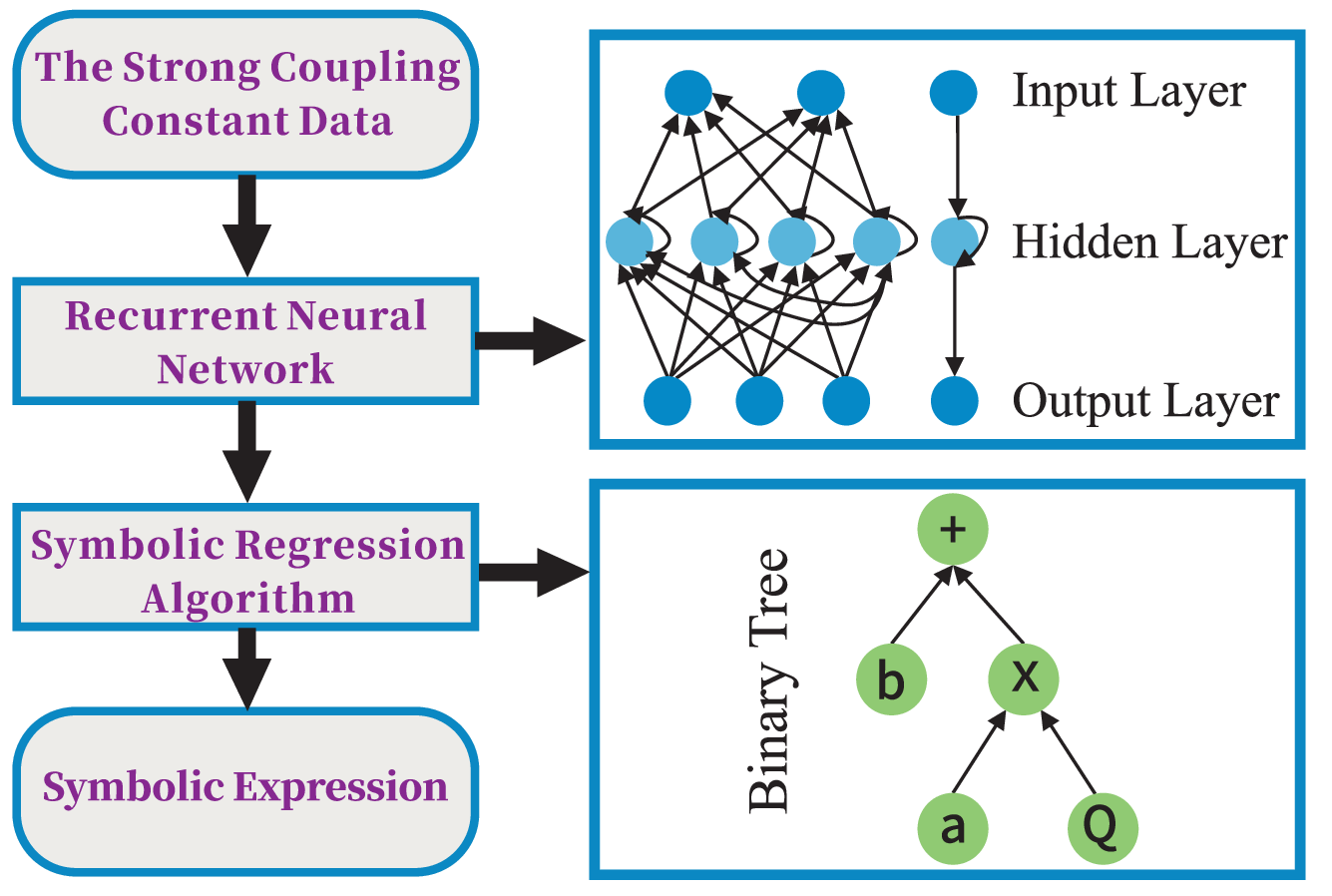}
	\caption{The algorithm framework schematic diagram. The coupling constant data is preprocessed and fed into a recurrent neural network, which effectively captures the underlying patterns. Subsequently, the symbolic regression algorithm utilizes this processed data to generate a precise and concise symbolic expression, enabling a straightforward representation of the relationships.}
	\label{fig:style}
\end{figure}

To achieve this purpose, we employ the $\Phi$-SO framework recently developed by a team at Strasbourg University \cite{Tenachi2023}. $\Phi$-SO framework can be used to search for analytic properties of noiseless data and to obtain analytical approximations of noisy data. As shown in Fig. \ref{fig:style}, the core of the framework is based on RNN with long-short term memory (LSTM) and symbolic regression algorithm. The advantage of $\Phi$-SO is that it can solve the ``black box'' generated by neural network training, which is a stumbling block for us to apprehend the results of neural network training further. In addition, the system can automatically simplify complex expressions so that the function discovered by symbolic regression can be easily understood and interpreted. With this framework, we divide the experimental data into high and low energy scales with $Q=6$ GeV boundary. Because perturbations and non-perturbations are not clearly distinguished by one energy, a transition region exists between several GeVs. It is certain that the area after $Q=3$ GeV \cite{Deur:2014qfa} is considered to be entirely asymptotically free perturbation QCD.

Multiple high-energy scale experimental data $Q\in(6,2000)$ GeV with strong regularity and high accuracy have been accumulated by CMS, JADE, and other researchers \cite{Begel:2022kwp,Bethke:2022cfc}. To train the analytical expression between $Q$ and $\alpha_s(Q)$ accurately, two parameters, $a$ and $b$, are introduced into the training process, drawing insights from various theoretical models. At present, the specific physical interpretations of $a$ and $b$ are not explicitly defined, allowing the algorithm to autonomously discover the relationships between the input and output variables. 
For the training of high-energy scale data, the algorithm undergoes $112$ epochs. It is observed that with increasing epochs, the ``Reward'', RMSE, and complexity metrics exhibit consistency, indicating that the accuracy has reached a saturation point. Furthermore, the structure of the function expression remains stable throughout this process. Table \ref{tab:table1} (High-energy scale) presents the final behavior of expression evolution, illustrating how the symbolic expressions change from simple to complex with each epoch. Utilizing this framework, the expression is obtained,
\begin{equation}
	\begin{aligned}
		\alpha_s^{high}(Q) =& \frac{1.212}{\ln(Q^2/(0.298)^2)}\\
		\approx& \frac{1}{\ln((11.261Q^2)^{\frac{1}{1.212}})}.\label{eq:2}
	\end{aligned}
\end{equation}
The ``Reward'' and RME are $0.836$ and $0.007$,  respectively.
Notably, the expression with the highest accuracy (Number $6$) bears a striking resemblance to the structure of the first-order renormalization equation. This demonstrates that machine learning is a viable and effective approach for investigating the strong coupling constant, facilitating more comprehensive research in this area. 

\begin{figure*}
	\makebox [\textwidth] {\hspace {0cm}
	\includegraphics[scale=0.46]{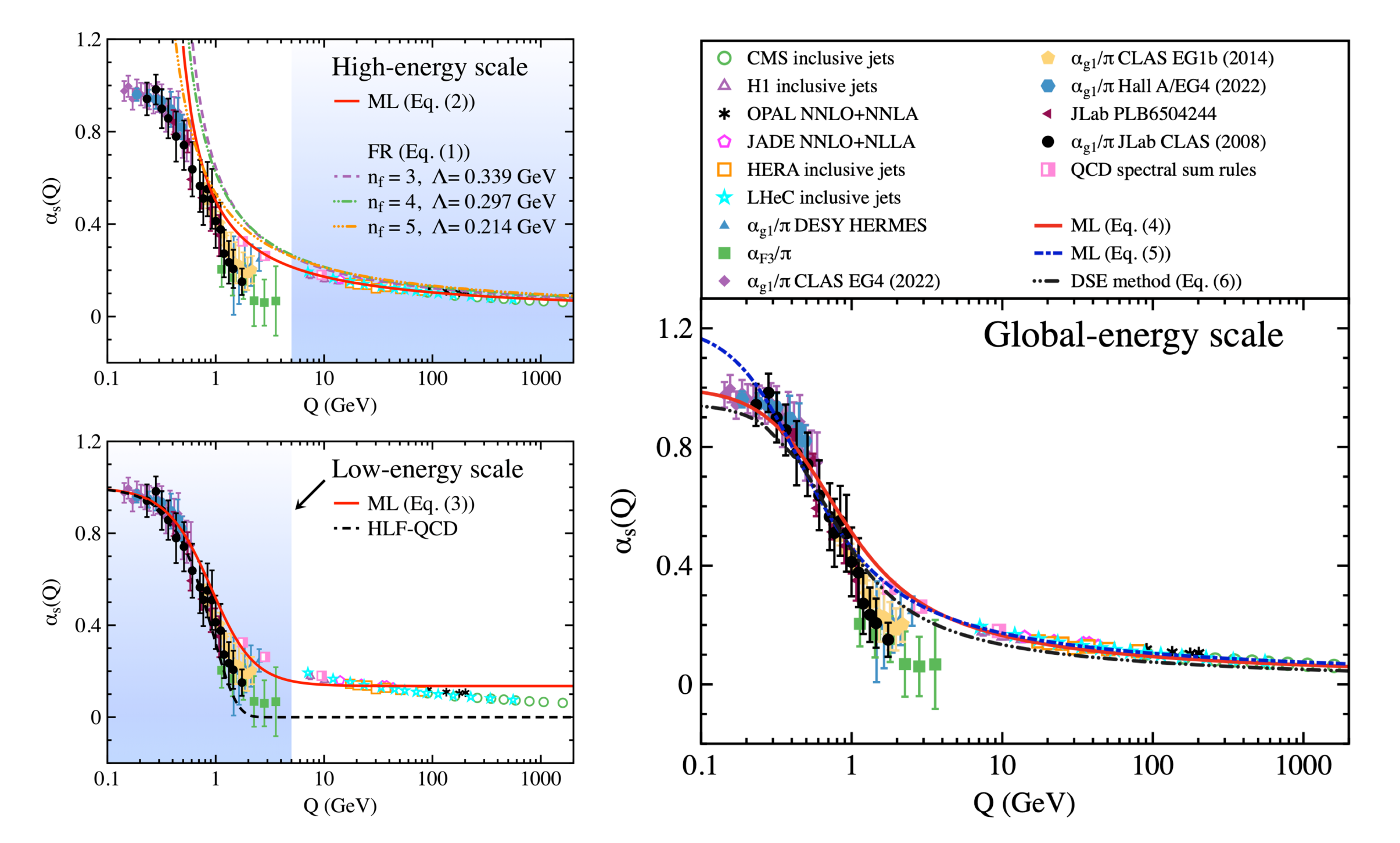}}
	\caption{The expression of high energy, low energy and global scale changes with $Q$. The top of left column plot displays the high-energy result, with $n_f=3, 4, 5$ taken from FLAG \cite{FlavourLatticeAveragingGroupFLAG:2021npn}. The bottom of left column plot corresponds to the low-energy scale and is derived from Holographic Light-Front QCD as presented in Ref. \cite{Brodsky:2010ur}. The right column plot represents the global-scale result, obtained from the DSE model (Eq. (\ref{eq:9})) as discussed in Ref. \cite{Fischer:2002hna}. The ML (Eq. (\ref{eq:11})) denotes the addition of a constant $e$ to the $\ln(f(Q))$ term in the denominator of the high-energy expression, allowing for a smooth transition from the perturbative to the non-perturbative region. Notably, it exhibits excellent agreement with experimental data. The left column's blue shadow indicates the range of data used for training purposes.}
	\label{fig:frame}
\end{figure*}

Upon examining the structure of Eq. (\ref{eq:2}), it becomes evident that it is consistent with the solution derived from the first-order renormalization equation, Eq. (\ref{eq:1}). The calculation of Flavour Lattice Averaging Group (FLAG) \cite{FlavourLatticeAveragingGroupFLAG:2021npn} in 2021 provide the $\Lambda$ values for different $n_f$. Take the quark flavor $n_f=3$ and the QCD scale parameter $\Lambda=0.339$ GeV, Eq. (\ref{eq:1}) can be expressed as $\alpha_s^{FR}(Q)	\approx 1/\ln((e+8.702Q^2)^{1/1.396})$.
The two expressions exhibit striking similarities, with only minor discrepancies in their parameters. ML, through its training and search capabilities, can autonomously achieve the theoretical model equation without requiring any manual intervention. This remarkable phenomenon is truly fascinating. As depicted in Fig. \ref{fig:frame} at the high-energy scale, our result aligns with the trend of the first-order renormalization equation and demonstrates improved agreement with experimental data.
The utilization of machine learning for determining the expression of the strong coupling constant has proven to be effective and reliable. This highlights the potential of employing this method for studying global scales. However, it is worth noting that as the momentum transfer $Q$ approaches zero, all current accepted results tend to infinity, resulting in an infinite value for $\alpha_s(0)$. This phenomenon is commonly referred to as infrared divergence.

At the low-energy scale, the training process is based on experimental data accumulated by JLab, CLAS, and other sources \cite{Deur:2008rf,Deur:2022msf,Deur:2016tte,Deur:2005cf,Deur:2008rf,Deur:2021klh,Kim:1998kia,HERMES:1997hjr,HERMES:1998pau,HERMES:1998cbu,HERMES:2002mes,HERMES:2006jyl,Narison:2018dcr,Braaten:1991qm,Yu:2021yvw}. It should be noted that the last three sets of $\alpha_{F3}/\pi$ data exhibit substantial errors, and in some cases, the values can even become negative when considering the error bars. To mitigate this issue, we assign reduced weights to these three data sets, minimizing their impact on the training results. After $2135$ epochs, a stable expression is obtained, as depicted in the Low-energy scale section of Table \ref{tab:table1}.
The training result with the ``Reward''=$0.809$ and RMSE=$0.050$,
\begin{equation}
	\alpha_s^{low}(Q)=e^{-\frac{2 Q^2}{b^2+Q^2}}, \label{eq:4}
\end{equation}
where $b=1.426$. 
In Fig. \ref{fig:frame} (Low-energy scale), the curve corresponding to this formula demonstrates a notable agreement with experimental data at the low-energy scale. Interestingly, we observe that the function $\alpha_{g_1}^{\mathrm{HLF}}\left(Q^2\right)/\pi= \mathrm{e}^{-Q^2 /\left(4 \kappa^2\right)}$ derived from HLF-QCD with $\kappa=M_{p}/2$ \cite{Brodsky:2010ur} exhibits a similar structure but appears to exhibit a weaker trend in Fig. \ref{fig:frame} (Low-energy scale).
However, both of these outcomes fall short in accurately describing the experimental data at the high-energy scale.

\begin{figure}[tbp]
	\centering
	\includegraphics[scale=0.40]{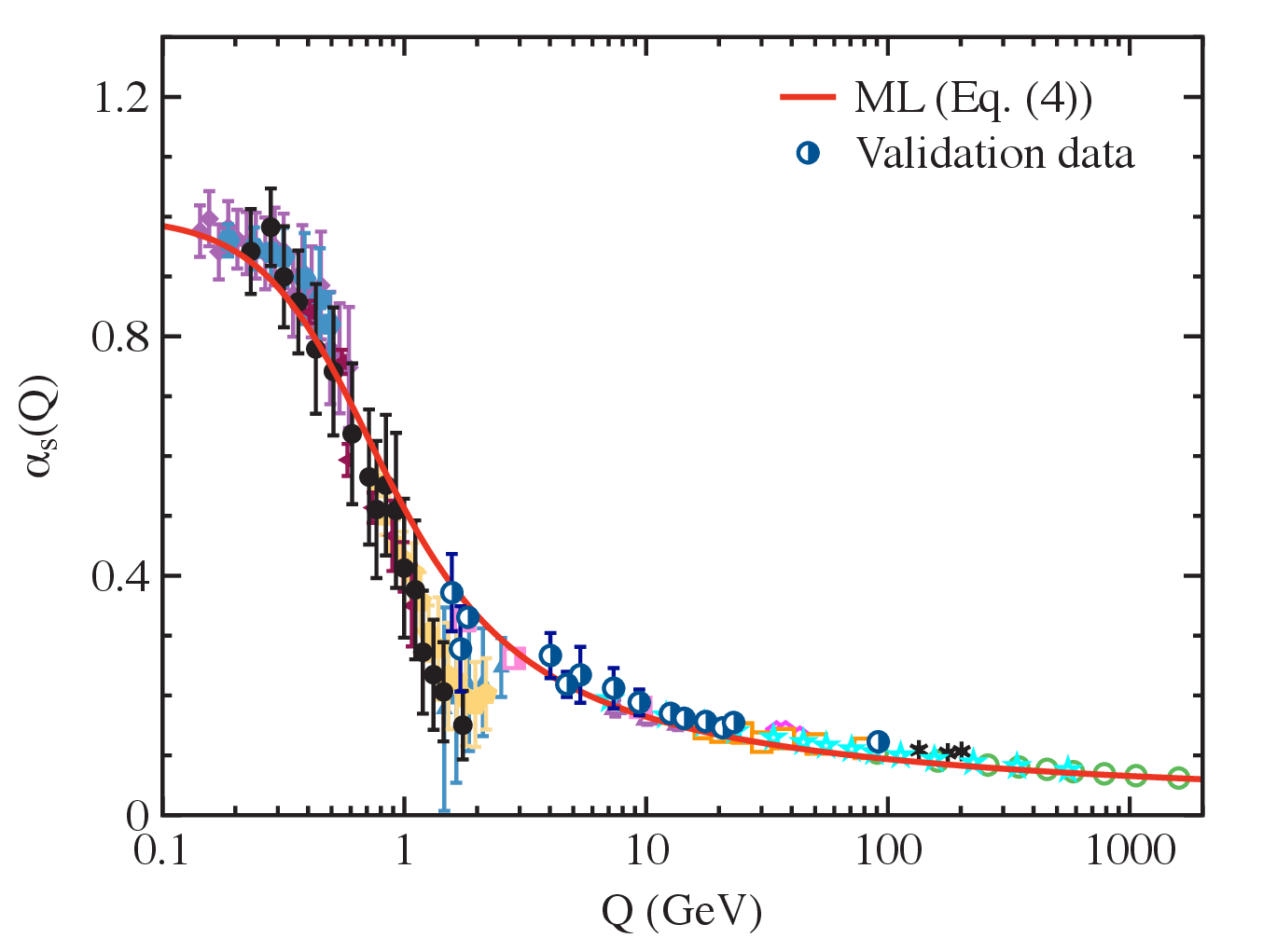}
	\caption{ Comparison of machine learning result at global-energy scale with validation data. The several sets of experimental data from $e^+e^-$ annihilation, deep inelastic scattering, heavy quarkonia \cite{Prosperi:2006hx}, BESIII data \cite{Shen:2023qgz}, CMS Collaboration \cite{CMS:2013vbb} and global scale comparison. Among them, our result is consistent with the $\alpha_s(M^2_Z)=0.1227^{+0.0117}_{-0.0132}$ recently extracted by BESIII \cite{Shen:2023qgz}.}
	\label{fig:val-data}
\end{figure}

Leveraging the insights gained from the aforementioned experiences, we utilize machine learning (ML) to train and analyze the experimental data across the global energy scale at $Q\in(0.1,2000)$ GeV, while excluding the three sets of $\alpha_{F3}/\pi$ data. After $2873$ epochs, we successfully obtain a symbolic expression capable of describing the strong coupling constant across the entire energy range for the first time. The evolutionary details of the expression are provided in Table \ref{tab:table1} Global-energy scale.
From detailed analysis, we identify the representation with ``Reward''=$0.800$ and RMSE=$0.053$,
\begin{equation}
	\begin{aligned}
		\alpha_s^{global}(Q)=&\frac{1}{\ln \left(\frac{Q^2}{(0.793)^2}+1\right)+1}
		= \frac{1}{\ln(e+e\frac{Q^2}{(0.793)^2})}\\
		\approx & \frac{1}{\ln(e+4.323 Q^2)}
		\label{eq:6},
	\end{aligned}
\end{equation}
which offers a better predictive behavior than the previous two formulas (Eq. (\ref{eq:2}) and (\ref{eq:4})).
Fig. \ref{fig:frame} (Global-energy scale) illustrates the behavior curve of the obtained result, which exhibits a notable agreement with the experimental data \cite{Deur:2008rf,Deur:2022msf,Deur:2016tte,Deur:2005cf,Deur:2008rf,Deur:2021klh,Kim:1998kia,HERMES:1997hjr,HERMES:1998pau,HERMES:1998cbu,HERMES:2002mes,HERMES:2006jyl,Narison:2018dcr,Braaten:1991qm,Yu:2021yvw}. Several sets of experimental data were measured through $e^+e^-$ annihilation, deep inelastic scattering, heavy quarkonia \cite{Prosperi:2006hx}, BESIII data \cite{Shen:2023qgz} and the CMS Collaboration \cite{CMS:2013vbb}, which were not considered and covered in all the previous training. Therefore, we adopt these data as validation data \cite{Prosperi:2006hx,CMS:2013vbb} to further prove the accuracy of Eq. (\ref{eq:6}), as shown in Fig. \ref{fig:val-data}. 

When $Q=0$ GeV, the divergence can be addressed by converting it into observable and finite physical quantities without the need for renormalization or factorization. Specifically, we find that $\alpha_s^{global}(0)=1/\ln(e)$.
Upon comparing with Eq. (\ref{eq:2}), we observe that only a constant $e$ is added to the $\ln(f(Q))$ term in the denominator. This addition of the constant $e$ allows for the description of the non-perturbative behavior by incorporating it into the function of the perturbative's $\ln(f(Q))$ term, where $Q$ is a function of $f(Q)$. Consequently, we strive to introduce the constant $e$ into the $\ln(f(Q))$ term in the denominator of Eq. (\ref{eq:2}),
\begin{equation}
	\begin{aligned}
		\alpha_s^{high^*}(Q) =& \frac{1.212}{\ln(Q^2/(0.298)^2+e)} \\
		\approx & \frac{1}{\ln((e+11.261Q^2)^{\frac{1}{1.212}})}.
	\end{aligned}\label{eq:11}
\end{equation}
Considering $Q$=$0$ GeV, we find that $\alpha_s^{high^*}(0)=1/(\ln(e^{1/1.212}))$, which is equivalent to a correction of $e$ in $\alpha_s^{global}(0)$. The behavior of Eq. (\ref{eq:11}) is depicted in Fig. \ref{fig:frame} (Global-energy scale). Remarkably, the overall trend closely resembles that of Eq. (\ref{eq:6}), providing further validation for the concept of transitioning from high-energy scale perturbative behavior to low-energy scale non-perturbative behavior.

A strong coupling constant formula derived from the DSE of Yang-Mills theory \cite{Fischer:2002hna},
\begin{equation}
	\begin{aligned}
		\frac{\alpha_s^{DSE}(Q)}{\pi}=&\frac{2.972}{\pi \ln \left(e+5.292 Q^{2.324}+0.034 Q^{3.169}\right)}\\
		\approx &  \frac{1}{\ln((e+5.292 Q^{2.324} + 0.034 Q^{3.169})^{\frac{1}{0.946}})}
		,\label{eq:9}	
	\end{aligned}
\end{equation}
which primarily demonstrates  the behavior of the strong coupling constant in the non-perturbative regime. One intriguing observation is the similarities we have discovered between their composition and our findings at the global energy scale. In Fig. \ref{fig:frame} (Global-energy scale), one observes that the shape of the DSE aligns with our results, however, it does not accurately capture the experimental data. Particularly, the high-energy scale appears lower than what is observed in the experiment.
When considering $Q=0$ GeV, we get $\alpha_s^{DSE}/\pi(Q=0)=1/\ln(e^{1/0.946})$, which can be interpreted as a correction term involving the constant $e$. This correction term accounts for the discrepancy observed in the high-energy scale region.

\begin{figure}[tbp]
	\centering
	\includegraphics[scale=0.40]{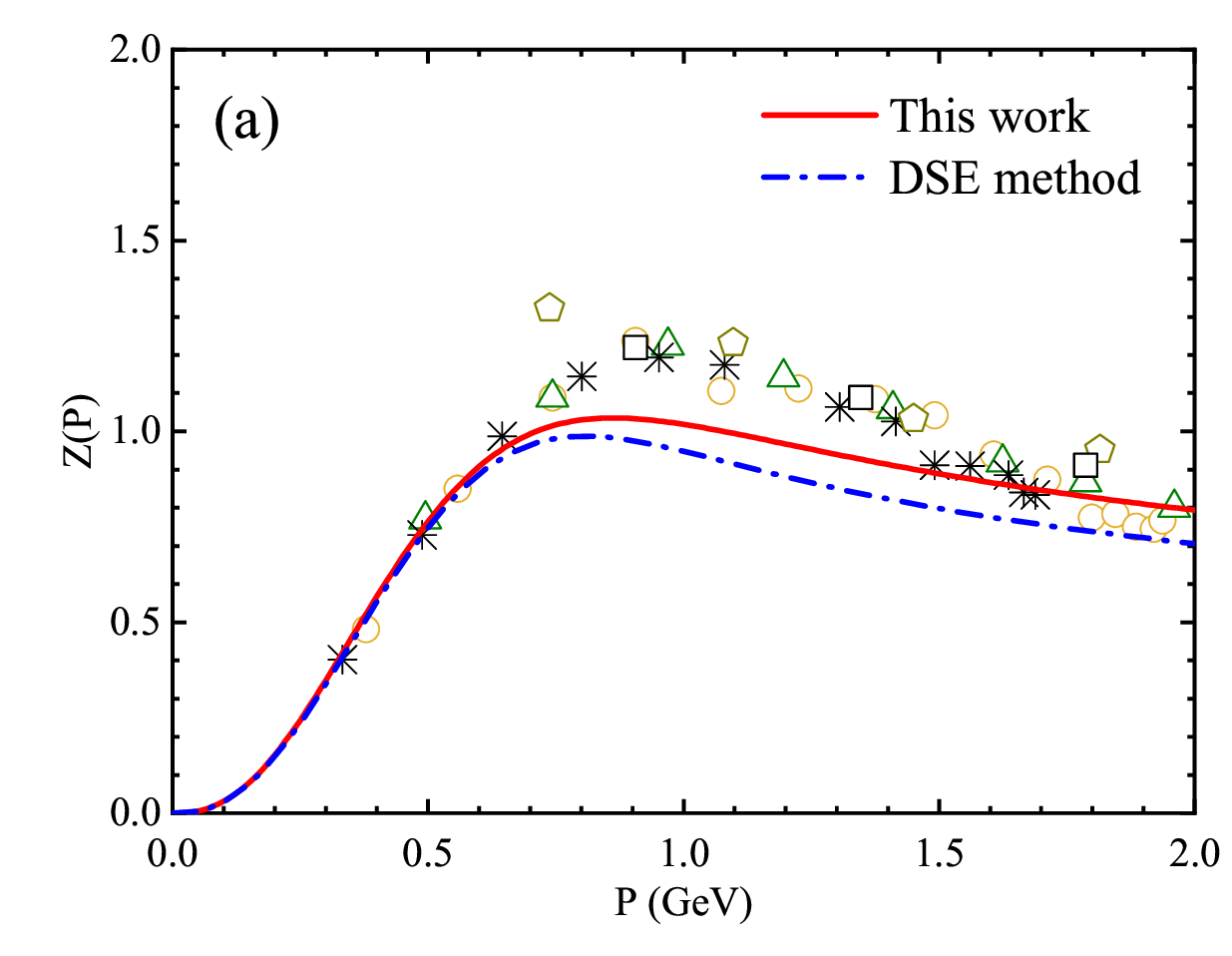}
    \includegraphics[scale=0.40]{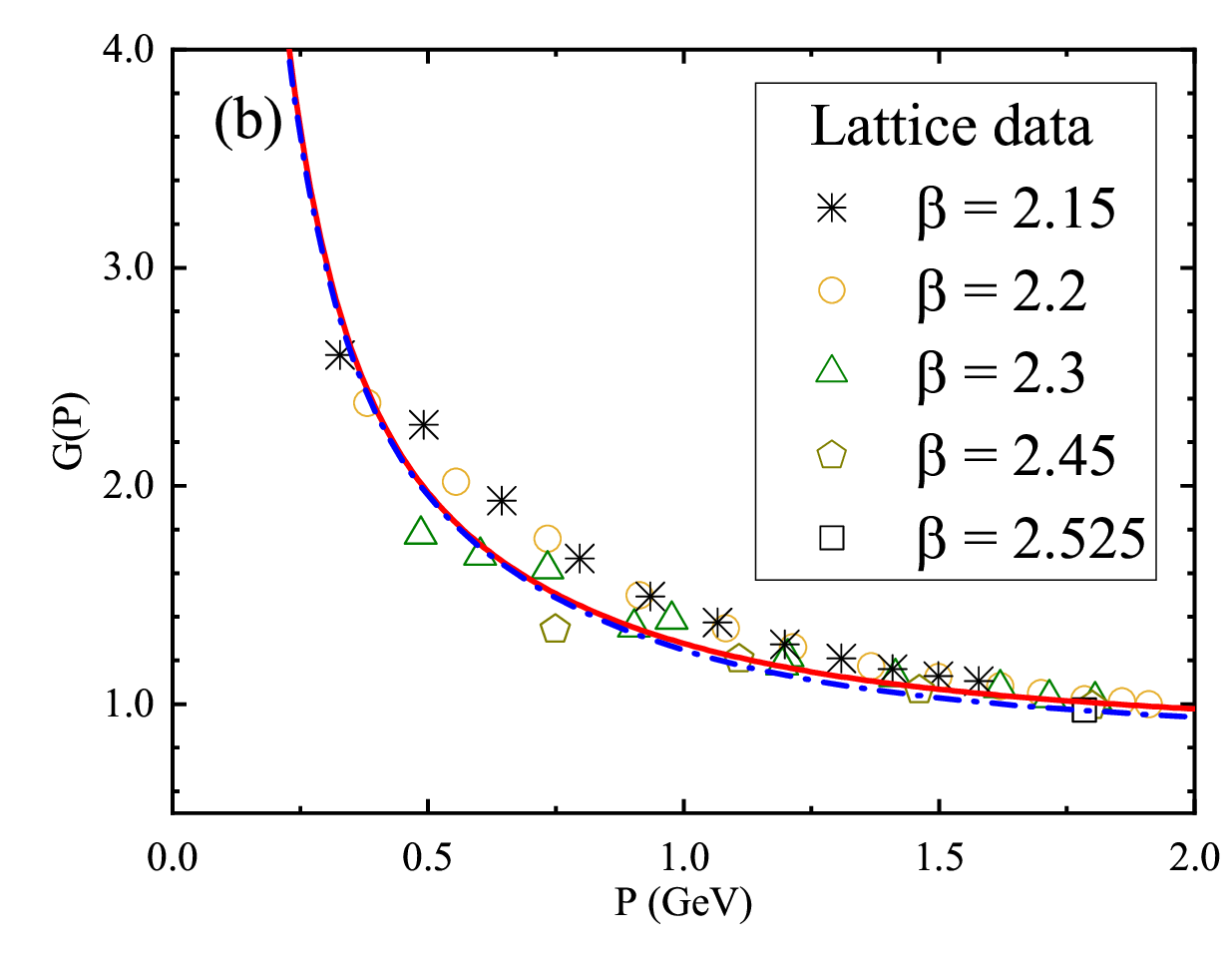}
	\caption{The distribution of gluon ($a$) and ghost ($b$) dressing functions based on the strong coupling constant $\alpha_s(Q)$, the various $\beta$ values correspond to the lattice data extracted from Ref. \cite{Fischer:2002hna}. The red line is the result of this work based on Eq. (\ref{eq:6}). }
	\label{fig:gluon}
\end{figure}

The non-perturbative structure of Yang-Mills theory involves two crucial field variables: ghost and gluon. These variables play a fundamental role in describing gauge invariance and the strong interaction, respectively. In the context of the Landau gauge, the ghost and gluon dressing functions represent the ratio of the ghost and gluon propagators to their corresponding free propagators \cite{Fischer:2002hna},
\begin{equation}
	G(x)=\left(\frac{\alpha(x)}{\alpha(\mu)}\right)^{-\delta} R^{-1}(x),\quad 
	Z(x)=\left(\frac{\alpha(x)}{\alpha(\mu)}\right)^{1+2 \delta} R^2(x) \label{eq:14}
\end{equation}
with  the auxiliary function $R(x)$
\begin{equation}
	R(x)=\frac{c x^\kappa+d x^{2 \kappa}}{1+c x^\kappa+d x^{2 \kappa}},
\end{equation}
which reflects that ghost and gluon are affected by strong coupling constants in the non-perturbative region. In this context, $\alpha(x)$ represents the strong coupling constant, where $x$ corresponds to the squared momentum, and the other relevant parameters are obtained from Ref. \cite{Fischer:2002hna}.
Figure \ref{fig:gluon} presents the gluon ( \ref{fig:gluon} $a$) and ghost (\ref{fig:gluon} $b$) dressing functions obtained from our results. The result of this work exhibits good agreement with the lattice data, indicating that our global $\alpha_s^{global}(Q)$ expression capture the non-perturbative characteristics of the Yang-Mills theory. Furthermore, the distribution describes the non-perturbative behavior of the ghost and gluon fields, which illustrates the interpretability of the physical properties of the machine learning results.

\begin{figure}[tbp]
	\centering
	\includegraphics[scale=0.40]{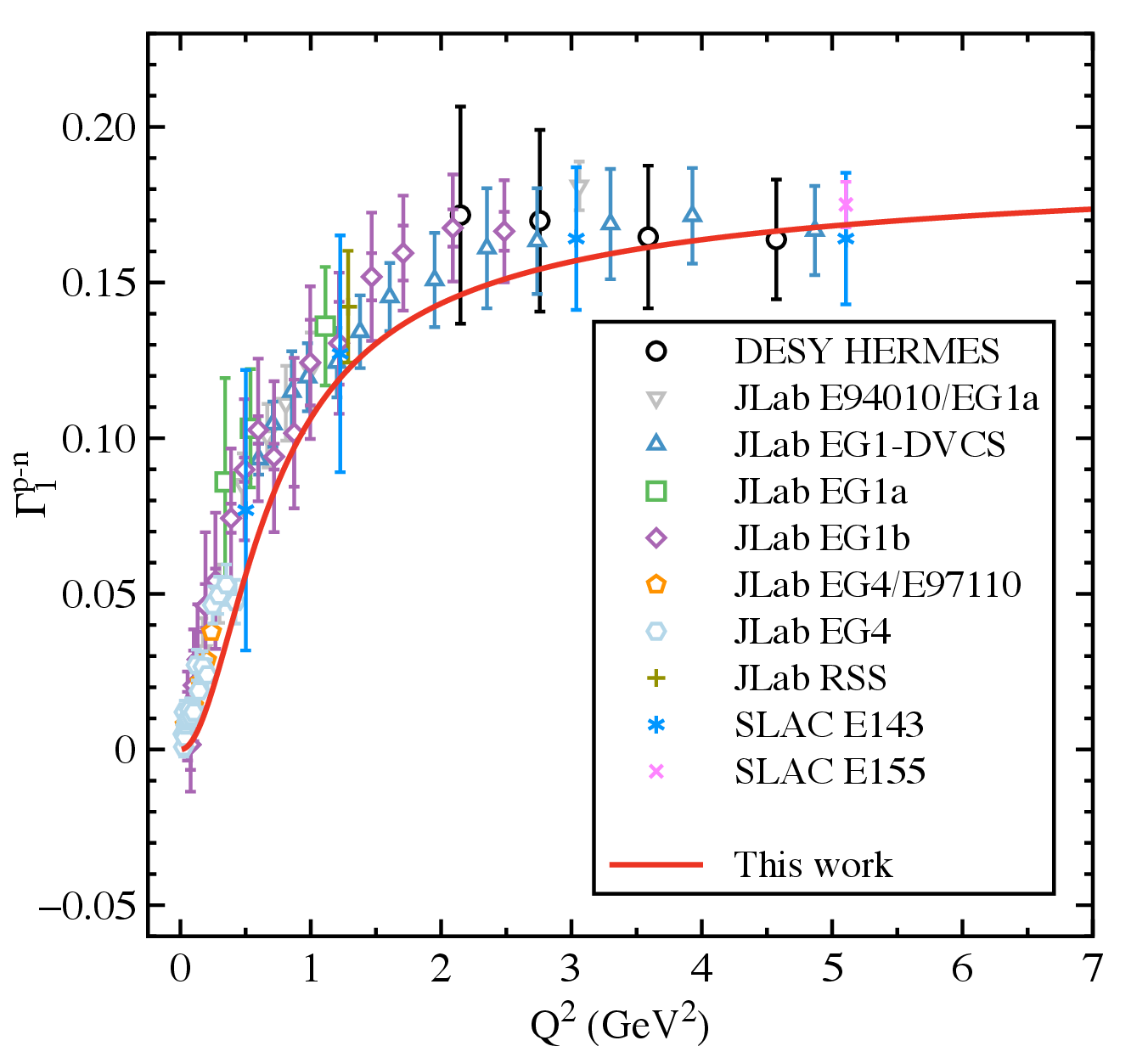}
	\caption{  Machine learning results of strong coupling constants for interpretation of nucleon spin structure. The experimental data are obtained from Refs. \cite{Deur:2023dzc,E143:1998hbs,Deur:2004ti,Deur:2008ej} and the red line is the BSR distribution of this work based on the Eq. (\ref{eq:6}).}
	\label{fig:biyueken}
\end{figure}
\section{Bjorken sum rule}
Bjorken sum rule (BSR) describes the spin distribution in the internal structure of protons and neutrons and is an essential prediction of quantum chromodynamics (QCD) \cite{Matsuda:1996np}. The BSR holds in the Bjorken scaling domain, which states that the difference integral of the spin structure functions of protons and neutrons equals the nucleon axial charge $g_A$ \cite{Matsuda:1996np,Ayala:2022mgz}. The Particle Data Group (PDG) has measured a more precise experimental value for $g_A$ at $1.2756\pm 0.0013$ \cite{ParticleDataGroup:2020ssz}.
The rule's significance lies in its ability to describe nucleon spin, aiding our comprehension of nucleon structure. In general, BSR is the integral of the isovector portion of the nucleon spin structure function $g_1(x,Q^2)$ \cite{Deur:2009zy},
\begin{equation}
\Gamma_1^{p-n}\left(Q^2\right)=\int_0^1 d x\left[g_1^p\left(x, Q^2\right)-g_1^n\left(x, Q^2\right)\right] \stackrel{\Lambda_s^2 / \underline{Q}^2 \simeq 0}{=} \frac{g_A}{6},
\end{equation}
$\Lambda_s^2 / Q^2 \simeq 0$ is valid here. Thus, there exists a correction for $Q^2$ outside of any neighborhood, resulting in a series of perturbation QCD,
\begin{equation}
\begin{aligned}
\Gamma_1^{p-n}(Q^2) &=\frac{g_A}{6} [1-\frac{\alpha_s^{pQCD(Q^2)}}{\pi}-3.58(\frac{\alpha_s^{pQCD(Q^2)}}{\pi})^2\\
&\quad-20.21(\frac{\alpha_s^{pQCD}(Q^2)}{\pi})^3-175.7(\frac{\alpha_s^{pQCD}(Q^2)}{\pi})^4\\
&\quad +\sim -893.38(\frac{\alpha_s^{pQCD}(Q^2)}{\pi})^5+\mathcal{O}\left(\left(\alpha_{\mathrm{s}}^{\mathrm{pQCD}}\right)^6\right)\\
&\quad+\sum_{n>1} \frac{\mu_{2 n}\left(Q^2\right)}{Q^{2 n-2}}
].
\end{aligned}
\end{equation}
This rule is valid because it is established on the DGLAP framework.
To implement the Grunberg scheme, this rule can be rewritten as,
\begin{equation}
    \Gamma_1^{p-n}=\frac{g_A}{6}(1-(\frac{\alpha_s(Q^2)}{\pi})).\label{eq:bsr}
\end{equation}
With this definition, BSR can be combined with the results of ML for further analysis. After simply processing the expression of global-energy scale Eq. (\ref{eq:6})  obtained based on ML and bringing it into Eq. (\ref{eq:bsr}), the change behavior between $\Gamma_1^{p-n}$ and $Q^2$ will be accepted, as shown in Fig. \ref{fig:biyueken}. The experimental data are measured by SLAC, JLab and other laboratories \cite{Deur:2023dzc,E143:1998hbs,Deur:2004ti,Deur:2008ej}. 
It is worth noting that the behavior obtained at the global scale well describes the $\Gamma_1^{p-n}$ change behavior between small $Q^2$ and large $Q^2$. This shows the relationship between nucleon spin and $Q^2$ and proves that the global-energy scale expression Eq. (\ref{eq:6}) obtained by ML can effectively describe the nucleon spin structure.
\section{The QCD static energy}
\begin{figure}[tbp]
	\centering
	\includegraphics[scale=0.40]{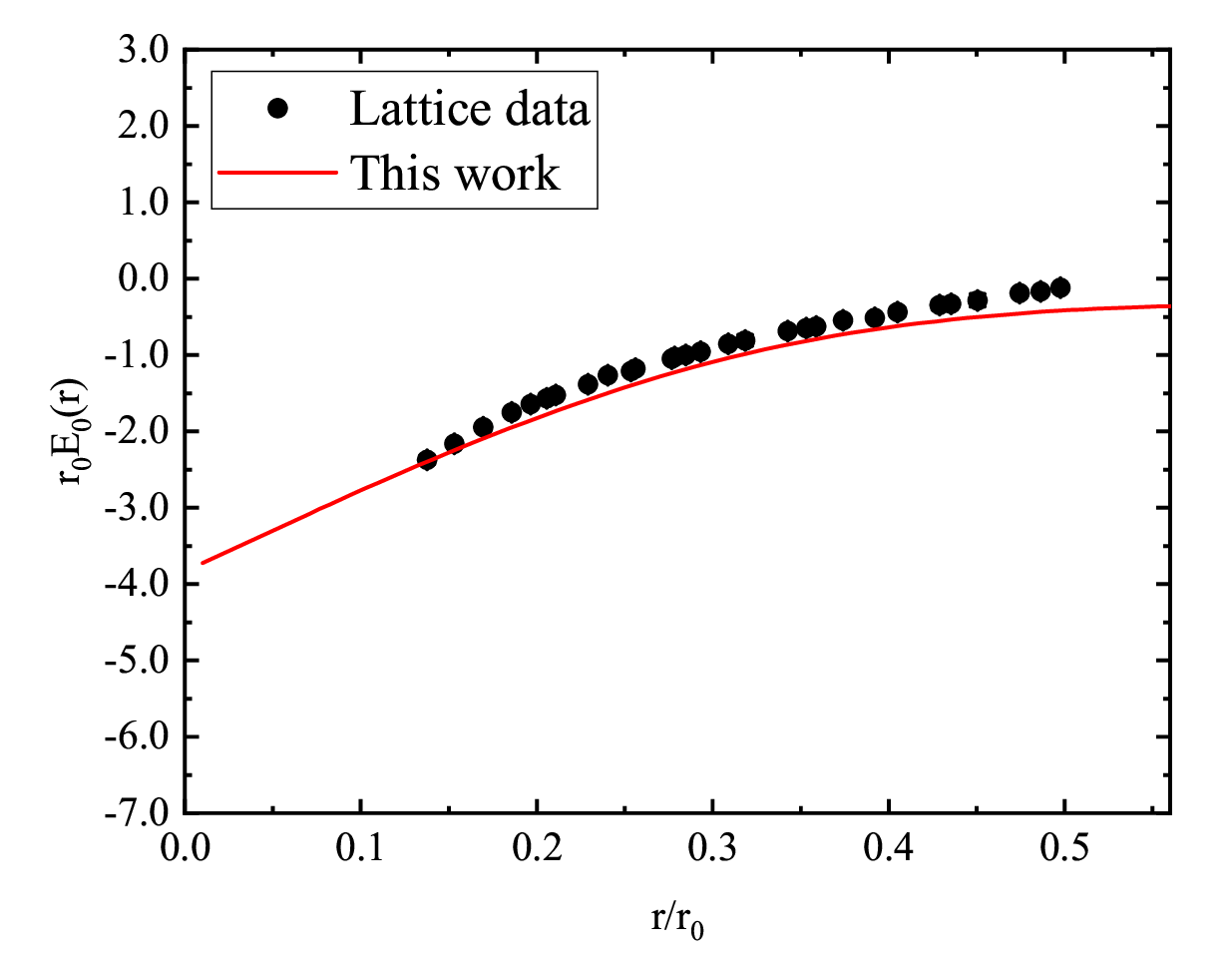}
	\caption{ The global expression generated by machine learning describes the fluctuating patterns of Lattice data \cite{Necco:2001xg} for QCD static energy, where $r_0$ is the constant derived from Ref. \cite{Bazavov:2019qoo}. The red line is the QCD static energy based on the Eq. (\ref{eq:6}).}
	\label{fig:static}
\end{figure}

A set of lattice static energy distribution $E_0(r)$ data at short distance at $n_f=2+1$ was recently published by HotQCD Collaboration \cite{Bazavov:2011nk}. Reference \cite{Tormo:2013tha} provides a detailed review of the known terms of perturbation of static energy, together with an analysis of the results and reasons for different loop descriptions of the lattice data. And the strong coupling constant of $\alpha_s(\rho=1.5\,{\rm GeV},\,n_f=3)$ is determined from the data. In this work, we studied the QCD static energy based on the global-energy scale expression of the ML reverse analysis and compared it with lattice data. For this purpose, we utilize the conclusions of Ref. \cite{Bazavov:2014soa} regarding the static energy $E_0(r)$ of perturbations,
\begin{equation}
    F(r)=\frac{dE_0}{dr},
\end{equation}
and integrate the form to obtain the energy. Here, one takes the quark-quark force to be \cite{Bazavov:2014soa},
\begin{equation}
\begin{aligned}
F(r)= & \frac{C_F}{r^2} \alpha_s(1 / r)\left[1+\frac{\alpha_s(1 / r)}{4 \pi}\left(\tilde{a}_1-2 \beta_0\right)\right. \\
& +\frac{\alpha_s^2(1 / r)}{(4 \pi)^2}\left(\tilde{a}_2-4 \tilde{a}_1 \beta_0-2 \beta_1\right) \\
& +\frac{\alpha_s^3(1 / r)}{(4 \pi)^3}\left(\tilde{a}_3-6 \tilde{a}_2 \beta_0-4 \tilde{a}_1 \beta_1-2 \beta_2\right. \\
& \left.\left.+a_3^L \ln \frac{C_A \alpha_s(1 / r)}{2}\right)+\mathcal{O}\left(\alpha_s^4, \alpha_s^4 \ln ^2 \alpha_s\right)\right],
\end{aligned} \label{eq:force}
\end{equation}
where $C_A=N_C$ is the number of colors. $\alpha_s(1 / r)$ is the Fourier transform form of the global-energy scale expression of ML. The fundamental representation's Casimir is denoted by $C_F$, and $\tilde{a}_1$, $\tilde{a}_2$, $\tilde{a}_3$, $a^L_3$, $\beta_0$, $\beta_1$ and $\beta_2$ are explained in detail in Ref. \cite{Bazavov:2019qoo}. By analyzing these, the distribution of the static energy at the short distances can be obtained, which is shown in Fig. \ref{fig:static}.
This reflects the change in static energy between quarks over short distances. The global-energy scale formula of ML in Fig. \ref{fig:static} better describes the variation behavior of the lattice data, indicating that the ML method effectively represents the static energy at short distances. The reliability and interpretability of the strong coupling constant of machine learning are further verified in reverse.
\section{Summary}
The strong coupling constant $\alpha_s(Q)$ is central to modern QCD and the Standard Model of particle physics. Since QCD theory was established in the last century, individuals have actively endeavored to understand it. At high energy scale $Q$, the model established by physicists can effectively describe $\alpha_s(Q)$. However, there still exist some uncertainties regarding the QCD scale parameter $\Lambda$. It is crucial to determine and analyze the $\Lambda$ value accurately. Based on the $\Phi$-SO framework, this work automatically searches for an analytical expression of the strong coupling constant varying with the energy scale through the various energy $Q$ experimental data. The exciting thing is that one obtains the expression at high-energy scale with the identical structure as the first-order renormalization equation. And the behavior of the global scale formula describes the change of the strong coupling constant measured experimentally at different energies.
Through multi-party verification of RMSE and ``Reward'', we believe the result by ML is valid and reliable. 

Through a comparison of the derived functions at both global and high-energy scales, we have made a noteworthy observation. The transition from the perturbative to non-perturbative regimes seems to necessitate the inclusion of a constant parameter, denoted as $e$. Remarkably, this parameter is autonomously learned by the ML algorithm or serves as a correction to the $\ln(f(Q))$ term. These findings provide valuable empirical data and theoretical support, which can prove pivotal for future investigations in this field.
Moreover, our analysis of the gluon and ghost dressing functions reveals intriguing patterns. The modified expressions obtained for both global and high-energy scales accurately capture the non-perturbative tendencies exhibited by these fundamental fields.

One investigates the $Q^2$ dependence of BSR based on the QCD strong coupling constant of ML, and adopt Grunberg scheme to study the relationship between $\alpha_s(Q)$ and $\Gamma_1^{p-n}$.
The results of the global-energy scale have been found to be in good agreement with experimental data, both in the large $Q^2$ region and the small $Q^2$ region. In addition, the results based on the global scale also have an excellent inverse verification of the behavior of the static energy and distance $r$ between quarks. These indicate that our method can effectively describe the relationship between the nucleon structure and the QCD strong coupling constant. This allows us to investigate the strong coupling constant and its resultant physical properties with machine learning methods. Further exploration of these interwoven properties will undoubtedly contribute to the construction of a more comprehensive and profound understanding of the fabric of the strong interaction.

\begin{acknowledgments}
		This work is supported by the National Natural Science Foundation of China under Grants No. 12065014, No. 12047501, No. 12247101 and No. 12335001, and by the Natural Science Foundation of Gansu province under Grant No. 22JR5RA266. We acknowledge the West Light Foundation of The Chinese Academy of Sciences, Grant No. 21JR7RA201.
X.L. is also supported by the China National Funds for Distinguished Young Scientists under Grant No. 11825503, National Key Research and Development Program of China under Contract No. 2020YFA0406400, the 111 Project under Grant No. B20063, the fundamental Research Funds for the Central Universities, and the project for top-notch innovative talents of Gansu province.
\end{acknowledgments}

\setcounter{equation}{0}
	\renewcommand{\theequation}{A.\arabic{equation}}
    \renewcommand{\theHequation}{A.\arabic{equation}}
    \setcounter{figure}{0}
	\renewcommand{\thefigure}{A\arabic{figure}}
    \renewcommand{\theHfigure}{A\arabic{figure}}
\appendix
\section*{Supplementary material : Analysis of strong coupling constant  with machine learning and its application}

We present a detailed account of the symbolic regression
process, specifically focusing on the evolution of expressions.
To ensure reliable and accurate training results, the experimental data from HERA, JADE, JLab, CLAS and other experiments \cite{Begel:2022kwp, Bethke:2022cfc,Deur:2008rf,Deur:2022msf,Deur:2016tte,Deur:2005cf,Deur:2008rf,Deur:2021klh,Kim:1998kia,HERMES:1997hjr,HERMES:1998pau,HERMES:1998cbu,HERMES:2002mes,HERMES:2006jyl,Narison:2018dcr,Braaten:1991qm,Yu:2021yvw} measuring strong coupling constants $\alpha_s(Q)$ at varying energy $Q$ are selected. In order to evaluate the accuracy and robustness of the trained model, the following parameters and evaluation indexes are introduced:
\begin{enumerate}[label=\textnormal{(\roman*)}]
	\item Complexity: The complexity of an expression can be evaluated based on the types and quantity of operators it encompasses.
	\item Length: The length of an expression is a measure of its size, indicating the total number of symbols present, including variables, constants, operators, and other elements.
	\item Reward: This index evaluates the symbolic expression generated by the framework and assesses its alignment with the desired data, taking into account both simplicity and complexity aspects.
	\item Root Mean Square Error (RMSE): This is a standard loss function,
	$
		\text { RMSE }=\sqrt{\sum\left(P_{\mathrm{i}}-O_{\mathrm{i}}\right)^2 / \mathrm{n}},
	$
	which represents the error between the predicted result and the real data value. A smaller value indicates a higher accuracy of the model. Here, $P_i$ is the predicted value, $O_i$ is the real value and $n$ is the total number size of samples in the data set.
\end{enumerate}

\begin{figure}[tbp]
	\centering
	\includegraphics[scale=0.35]{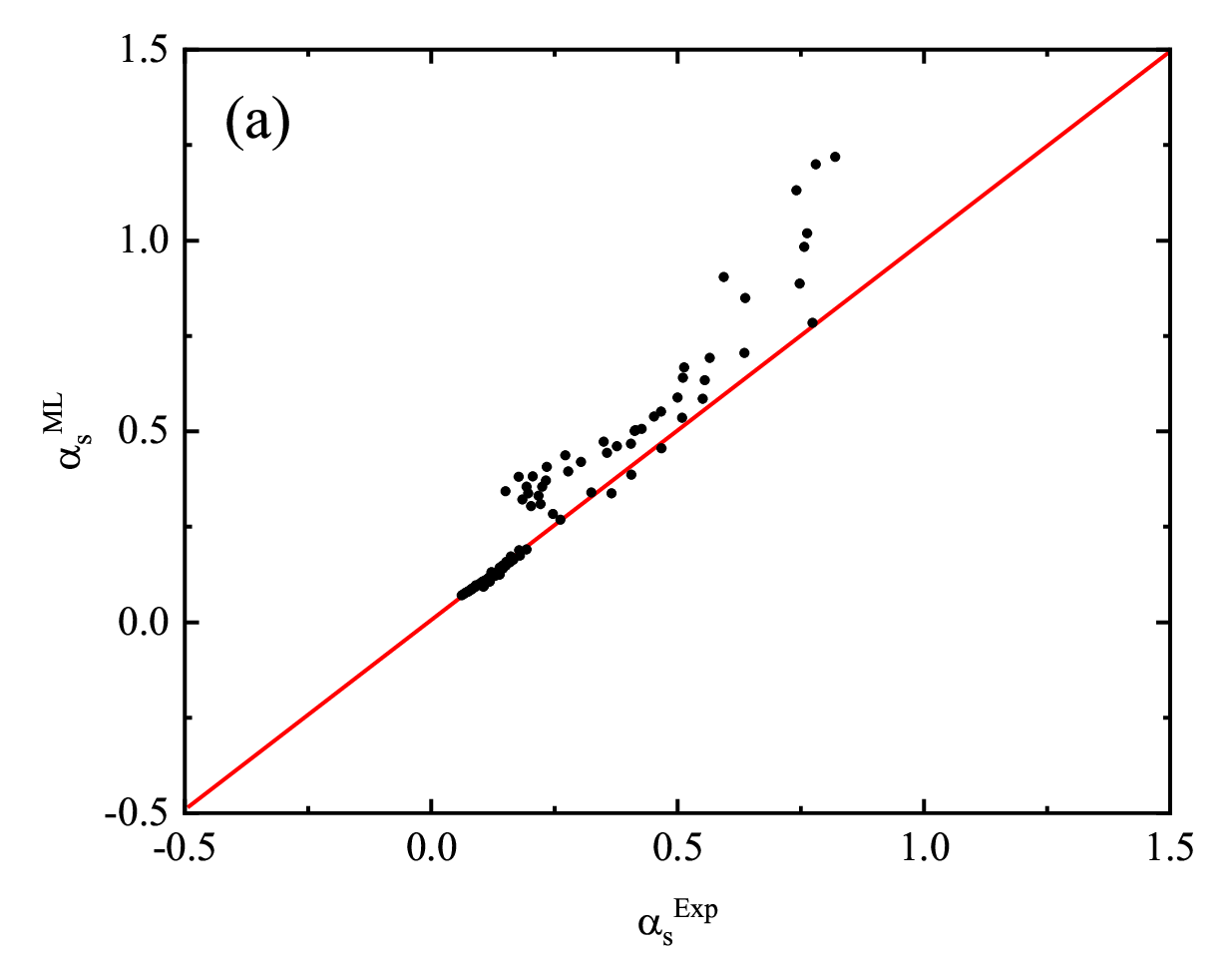}
 \includegraphics[scale=0.35]{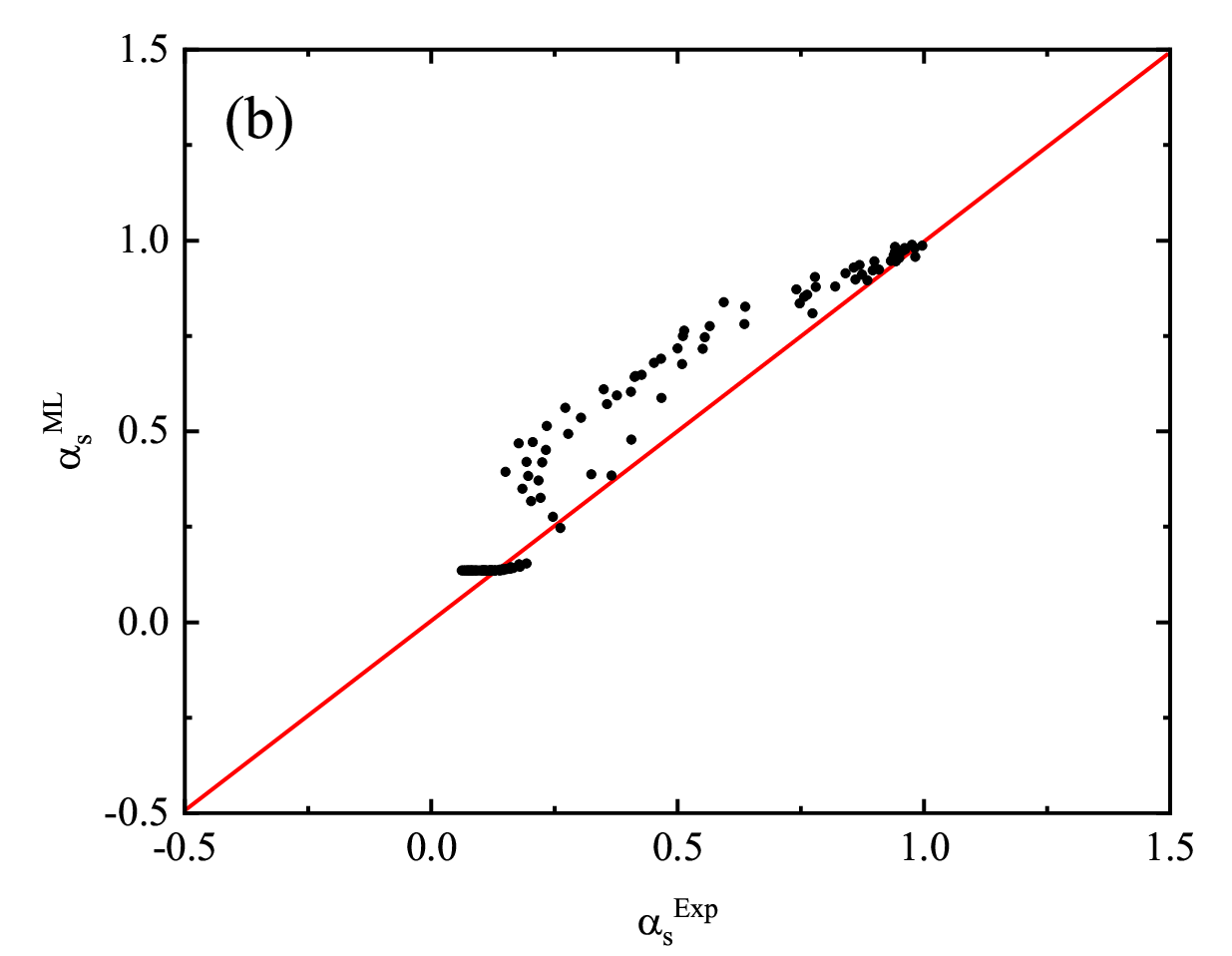}
  \includegraphics[scale=0.35]{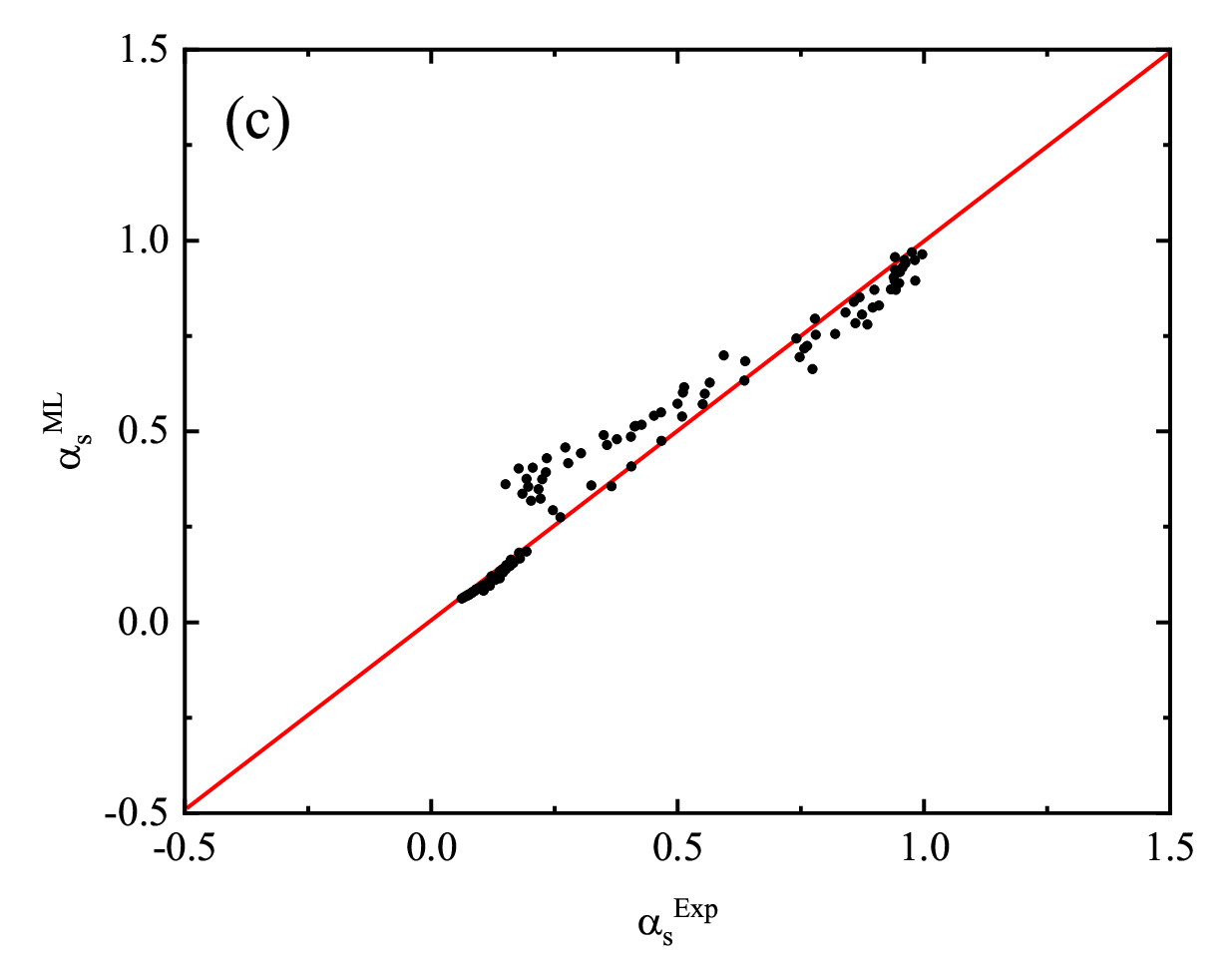}
	\caption{Benchmarking machine learning results with experimental data. $a$, $b$ and $c$ are the comparison results of high, low, and global scales, respectively. The vertical axis is the result of machine learning, and the horizontal axis is the experimental data.}
	\label{fig:pre}
\end{figure}

In this work, the experimental data are divided into two categories based on the energy scale: high-energy scale $Q\in(6,2000)$ GeV and low-energy scale $Q\in(0.1,6)$ GeV. For the training of high-energy scale data, the algorithm undergoes $112$ epochs. It is observed that with increasing epochs, the Reward, RMSE, and complexity metrics exhibit consistency, indicating that the accuracy has reached a saturation point. Furthermore, the structure of the function expression remains stable throughout this process. Table \ref{tab:table1} (High-energy scale) presents the final behavior of expression evolution, illustrating how the symbolic expressions change from simple to complex with each epoch. Notably, the expression with the highest accuracy (Number $6$) bears a striking resemblance to the structure of the first-order renormalization equation. This demonstrates that machine learning is a viable and effective approach for investigating the strong coupling constant, facilitating more comprehensive research in this area.

At the low-energy scale, the last three sets of $\alpha_{F3}/\pi$ data exhibit substantial errors, and in some cases, the values can even become negative when considering the error bars. To mitigate this issue, we assign reduced weights to these three data sets, minimizing their impact on the training results. After $2135$ epochs, a stable expression is obtained, as depicted in the Low-energy scale section of Table \ref{tab:table1}.

Next, we proceed to train the global-energy data, spanning $Q\in(0.1,2000)$ GeV, while excluding the three sets of $\alpha_{F3}/\pi$ data. After $2873$ epochs, we successfully obtain a symbolic expression capable of describing the strong coupling constant across the entire energy range for the first time. The evolutionary details of the expression are provided in Table \ref{tab:table1} Global-energy scale.

Upon examining the complete set of epochs, it is noteworthy that the highest complexity level attained is only $13$, indicating a relatively moderate degree of nonlinearity and function complexity within an acceptable range. As previously mentioned, the expression generated through ML training achieves the highest level of accuracy, closely replicating the behavior of the experimental data. Therefore, for predictions and detailed analysis, we select the expressions with the highest accuracy from these three sections.


In order to compare the difference between real data and predicted values, we analyzed the results of training at different energy scales, as shown in Fig. \ref{fig:pre}. 
It presents the comparison between the predicted values by Eq. (\ref{eq:2}) after training with high-energy scale data and the experimental values, which indicates that the strong coupling constant in the high-energy region exhibits a reasonable agreement with the experimental values. In contrast, the effect in the low-energy region is worse.
Further, exclusively low-energy scale data is used for training, and the obtained function is Eq. (\ref{eq:4}). 
The comparison of the predicted strong coupling constant with the experimental value is shown in Fig. \ref{fig:pre} (b), which displays the opposite result as shown in Fig. \ref{fig:pre} (a). These analyses indicate that training with data from a specific range cannot capture the change of the strong coupling constant at the global energy scale. Subsequently, the global-scale data is used for training, and the comparison between the predicted results and the true values is shown in Fig. \ref{fig:pre} (c). This consistency demonstrates the feasibility and inevitability of using global energy data for training.

\begin{table*}[h]\small
	\renewcommand\arraystretch{2}
	\caption{\label{tab:table1} Learning the evolution of complexity, ``Reward'', RMSE and expression of the strong coupling constant based on the ML. The complexity of the expressions gradually increases from simple to complex, while the "Reward" metric shows a gradual convergence towards 1. This indicates that as the training progresses, the expressions become more intricate and the model achieves a higher level of accuracy in capturing the underlying patterns in the data.}
	\begin{ruledtabular}
		\begin{tabular}{c|cccccccc}
			&Number  &  Complexity & Length & Reward & RMSE & Expression &  $a$ & $b$ \\
			\hline
			High-energy scale
			&$1$ & $4$ & $4$ & $0.475$ & $0.037$ & $\alpha_s(Q)=\frac{1}{e^2}$ & - & - \\			
			&$2$ & $5$ & $4$ & $0.503$ & $0.033$ & $\alpha_s(Q)=e^{a b}$ & $-0.720$ & $2.957$ \\			
			&$3$ & $6$ & $6$ & $0.707$ & $0.014$ & $\alpha_s(Q)=\frac{1}{\ln(Q/b)}$ & - & $0.016$ \\			
			&$4$ & $7$ & $7$ & $0.789$ & $0.009$ & $\alpha_s(Q)=\sqrt{-\frac{1}{\sqrt{Q/b}}}$ & - & $0.010$ \\			
			&$5$ & $9$ & $8$ & $0.790$ & $0.009$ & $\alpha_s(Q)=(\frac{1}{\sqrt{a Q^2}})^2$ & $1.642$ & - \\
			&$6$ & $12$ & $10$ & $0.836$ & $0.007$ & $\alpha_s(Q)=\frac{b}{a\ln(Q^2/a^2)}$ & $0.298$ & $0.361$  \\
			\hline
			Low-energy scale
			&$1$ & $4$ & $4$ & $0.599$ & $0.141$ & $\alpha_s(Q)=\cos(\frac{Q}{b})$ & -& $1.288$ \\
			&$2$ & $5$ & $5$ & $0.756$ & $0.068$ & $\alpha_s(Q)=\sin(\sqrt{\frac{b}{Q}})$ & -& $0.478$ \\
			&$3$ & $6$ & $6$ & $0.790$ & $0.056$ & $\alpha_s(Q)=1-e^{\frac{b}{Q}}$ & -& $-0.992$ \\
			&$4$ & $7$ & $7$ & $0.790$ & $0.056$ & $\alpha_s(Q)=1-e^{-\frac{b}{Q}}$ & - & $0.992$ \\
			&$5$ & $8$ & $8$ & $0.795$ & $0.054$ & $\alpha_s(Q)=e^{\cos(\frac{Q}{b})-1}$ & - & $1.057$ \\
			&$6$ & $10$ & $10$ & $0.796$ & $0.054$ & $\alpha_s(Q)=e^{-\frac{Q}{b}\sin(\frac{Q}{b})}$ & - & $1.453$ \\
			&$7$ & $11$ & $11$ & $0.796$ & $0.054$ & $\alpha_s(Q)=e^{-\ln(\frac{Q+b}{b})^2}$ & - & $1.078$ \\
			&$8$ & $13$ & $13$ & $0.809$ & $0.050$ & $\alpha_s(Q)=e^{-\frac{2Q^2}{b^2+Q^2}}$ & - & $1.426$ \\
			\hline
			Global-energy scale
			&$1$ & $4$ & $4$ & $0.401$ & $0.373$ & $\alpha_s(Q)=\frac{1}{2}$ & -& $-$ \\
			&$2$ & $6$ & $4$ & $0.661$ & $0.124$ & $\alpha_s(Q)=\sqrt{\frac{b}{Q}}$ & -& $0.268$ \\
			
			&$3$ & $7$ & $5$ & $0.693$ & $0.105$ & $\alpha_s(Q)=1-e^{-\frac{Q}{b}}$ & -& $2.486$ \\
			&$4$ & $8$ & $6$ & $0.719$ & $0.090$ & $\alpha_s(Q)=\frac{1}{a Q+1}$ & $0.478$ & - \\
			&$5$ & $9$ & $7$ & $0.732$ & $0.083$ & $\alpha_s(Q)=\sqrt{\ln(\frac{Q}{b+Q})}$ & - & $0.437$ \\
			&$6$ & $10$ & $6$ & $0.736$ & $0.081$ & $\alpha_s(Q)=\frac{1}{a(b+Q)}$ & $0.648$ & $1.356$ \\
			&$7$ & $11$ & $9$ & $0.763$ & $0.068$ & $\alpha_s(Q)=(\frac{1}{(\frac{Q}{b})^2+1})^{1/4}$ & - & $0.465$ \\
			&$8$ & $12$ & $10$ & $0.800$ & $0.053$ & $\alpha_s(Q)=\frac{1}{\ln(\frac{Q^2}{a^2}+1)+1}$ & $0.793$ & -
		\end{tabular}
	\end{ruledtabular}
\end{table*}

\end{document}